\documentclass[epj]{svjour}
\usepackage{graphics}
\usepackage{graphicx}
\usepackage{amsmath}

\def\olra{\stackrel{\leftrightarrow}}

\begin{document}

\title{Axial-tensor Meson Family at $T\neq0$ }
\author{J.Y.~S\"ung\"u, \inst{1} ~A. T\"{u}rkan, \inst{2} ~E. Sertbakan, \inst{1} E. Veli Veliev \inst{1}}
\authorrunning{J.Y.~S\"ung\"u \and A. T\"{u}rkan \and E. Sertbakan \and E. Veli Veliev}
\institute{Department of Physics, Kocaeli University, 41001 Izmit,
Turkey\and \"{O}zye\u{g}in University, Department of Natural and
Mathematical Sciences, \c{C}ekmek\"{o}y, Istanbul, Turkey\\}
\date{Received: date / Revised version: date}
\abstract{The mass and decay constants of $\rho_2,~\omega_2$ and a
missing member in the $ 2^{--} $ nonet along with their first
excited states are analyzed by the Thermal QCD sum rules approach,
including QCD condensates up to dimension five.  Mass and decay
constant values of these mesons are stable from $T=0$ up to ~$T \cong 120 ~\mathrm{MeV}$. However,
after this threshold point, our numerical analyses indicates that
they begin to diminish with increasing temperature. When we
compare the hadronic parameters with their vacuum values, masses
of these mesons and their first excited states decrease between $(1-13\%)$ from the PDG data and $(10-26\%)$
for the decay constants. However they diminish in the interval of
$(9-26\%)$ and $(2-34\%)$ respectively with regards to Regge
Trajectory Model data. We expect  our numerical results will be
confirmed by future heavy-ion collision experiments.} \PACS{
{14.40.n}{Mesons} \and {11.55.Hx}{Sum rules} \and {11.10.Wx}
{Finite-temperature field theory} }

\maketitle

\section{Introduction}

The physics of strongly interacting matter under extreme conditions is a major challenge in the Thermal QCD  \cite{Hatsuda:1992bv,Ayala:2016vnt,Mallik:1997kj,Dominguez:2016roi}.  It is predicted that bound quarks and gluons at high temperatures and/or densities liberate from
hadrons to form a new state of matter called as Quark-Gluon
Plasma (QGP). In this hot hadronic matter, a chiral phase transition is estimated to occur at a certain temperature. However, a quantitative explanation of (de)confinement and restoration (or breaking) of the chiral
symmetry phenomena is still lacking and reveals a research topic for future studies. The phase structure of the QGP contains rich information on strong interactions between quarks and gluons in hot medium and may shed light on some central questions
like confinement mechanisms, hadronisation, QCD vacuum dynamics, the nature of compact stars, and the evolution of
matter in the early universe \cite{Yagi:2005yb}.

Conditions similar to the early universe can be recreated in
the laboratory conditions in large-scale ultrarelativistic heavy-ion collision experiments and the data obtained is crucial
for modeling the hot-dense matter~\cite{Bazavov:2014pvz}. Searching the form of QCD phase diagram and fixing the region of phase transition from the hadronic matter to the QGP state at high temperatures are the
main objectives of current and planned experimental programs at
the RHIC in Brookhaven National Laboratory and future
experiments at the FAIR facility in Darmstadt and NICA in Dubna
\cite{Dong:2018fhv,Bugaev:2018lfj,Ablyazimov:2017guv}.

The precise experimental verification of the phase transition  temperature, called the critical temperature $T_c$, from hadronic matter to the QGP state would be a big step in improving
understanding of QCD in hot medium, and a significant contribution to the survey of QCD phase diagram. In addition, some studies assumed that there is a specific starting point for the QGP phase transition named pseudocritical temperature which is not a real phase transition, but an analytic crossover with a rapid change, as opposed to a jump~\cite{Steinbrecher:2018phh,Aoki:2006we,Cheng:2006qk}.
Recently, the critical temperature for the QGP formation has estimated at
$T_c\cong155~\mathrm{MeV}$~\cite{Andronic:2017pug} based on analysis of
experimental data from heavy-ion collisions at LHC and RHIC \cite{Steinbrecher:2018phh,Bazavov:2017dus},
although in UrQMD hybrid model it is proposed that the phase
transition temperature for hot matter should be between $160-165$
MeV \cite{Becattini:2012xb}. Some Lattice theory studies predict that critical temperature for the QGP phase transition is above this temperature \cite{Borsanyi:2012ve,Boyd:1996}. Therefore there is no unique temperature estimate for the deconfinement phase transition of hot matter.

In this manner there are many studies about the effect of temperature on the fundamental parameters of hadrons in the literature ~\cite{Mallik:1997kj,Dominguez:2016roi,Hohler:2013eba,Turkan:2019anj}. Due to the temperature dependence of the color screening radius in the QGP, it is expected that mesons with different flavors melt at certain temperatures. Light flavored mesons may dissociate in the neighborhood of $T_c$ reflecting the close relationship between the chiral crossover and deconfinement temperature~\cite{Hohler:2013eba,Bazavov:2014yba,Wang:2013wk}. So the chiral symmetry breaking point in hot medium can be described by the relevant thermal properties of the light mesons. Appearance of a turning point in temperature dependence of hadronic parameters will explain the occurrence of chiral symmetry transition. Additionally, the deviation of light mesons thermal mass from their mass in vacuum is closely connected to the location of freeze-out (i.e. hadronisation) temperature. In this sense, the light unflavored axial-tensors and their first excited states are of particular interest to provide valuable information on the formation of QGP.

Specifically, attention is shifted towards the light unflavored axial-tensor family with quantum number
$J^{PC} = 2^{--}$ to complete the hadron
spectrum which still needs to be properly classified
\cite{Turkan:2019anj,Guo:2019wpx,Ebert:2009ub,Godfrey:1998pd,Aliev:2017apq,Chen:2011qu,Pang:2017dlw}. However, there is a discrepancy on the ground and the first excited states of axial-tensor meson nonet between the Regge Trajectory Model's estimations and the data of PDG.
Our plan is to investigate which of these data is more consistent with the QCD sum rules (QCDSR) calculations. This is our other motivation for examining this family, whose main features are presented in Table \ref{tab:1}, in terms of PDG data.\\
\begin{table}[h!]
\begin{center}
\caption{Zero temperature mass and width values of light unflavored meson family in $ 2^{--} $ nonet.} \label{tab:1}
\begin{tabular}{lll}
\hline\hline
State & Mass~(MeV)~\cite{Zyla} & Width~(MeV)~\cite{Zyla}\\
\hline
$\rho_2$    & $1940 \pm 40$ & $155 \pm 40$ \\
$\omega_2$  & $1975 \pm 20$ & $175\pm 25$ \\
$\rho^*_2$   & $2225 \pm35$  & $335^{+100}_{-50}$ \\
$\omega^*_2$ & $2195 \pm 30$ & $225\pm 40$ \\
$\phi_2$   & $? $  & $?$ \\
$\phi^*_2$ & $?$ & $?$ \\
\hline\hline
\end{tabular}
\end{center}
\end{table}\\
The $\rho_2$ meson quark content is given as
$[u\bar{d}]$ in PDG. The physical
isoscalars $\omega_2$, $\phi_2$ are mixtures of the
$\mathrm{SU(3)}$ wave function $\psi_8$ and $\psi_1$:
\begin{eqnarray}
\omega &=& \psi_8 sin\theta+\psi_1 cos  \theta, \nonumber\\
\phi&=& \psi_8 cos \theta-\psi_1  sin\theta,
\end{eqnarray}
where $\theta$ is the nonet mixing angle, the physical
$\omega_2$ and $\phi_2$ states are the linear combinations of
these $\mathrm{SU(3)}$ singlet and octet states \cite{Zyla}:
\begin{eqnarray}
\psi_1&=&\frac{1}{\sqrt{3}}(u\bar{u} + d\bar{d} +s\bar{s}), \nonumber\\
\psi_8 &=& \frac{1}{\sqrt{6}}(u\bar{u} + d\bar{d} - 2s\bar{s}).
\end{eqnarray}
Due to the relatively small effect of the mixing angle, we can omit
the mixing of singlet and octet states since this is within
the uncertainties of the QCDSR approach. Namely, the $\omega_2$
and $\phi_2$ mesons can be handled as pure singlet and octet state,
respectively.

In this study, in addition to the above-mentioned objectives, we aimed to determine the behavior of mass and decay
constants of ground states of $\rho_2, \omega_2$, $\phi_2$ family and their first excited states in hot medium.
Assuming the quark-hadron duality is also valid at finite temperatures, we replace the vacuum expectation
values of the condensates and other related parameters with their temperature dependent expressions \cite{Mallik:1997kj}. We used the modified QCDSR theory up to dimension five which is typical since dimension six operators are constrained with no lattice data available. 

We organize rest of the content as follows: in section 2, we
talk about the theory used in our calculations. Then we
estimate the hadronic parameters of these states in hot medium and
present numerical analysis in section 3. Finally, we give a
summary and interpret the results in section 4.

%%%%%%%%%%%%%%%%%%%%%%%%%%%%%%%%%%%%%%%%%%%%%%%%%%%%%%%%%%%%%%%%%%%%%%%%%%%
\section{Thermal QCD sum rules}
%%%%%%%%%%%%%%%%%%%%%%%%%%%%%%%%%%%%%%%%%%%%%%%%%%%%%%%%%%%%%%%%%%%%%%%%%%%
One of the non-perturbative techniques used to investigate
chiral phase transition via analyzing the variations of hadronic properties at finite temperatures is the Thermal QCD sum rules (TQCDSR) approach. TQCDSR is the extended version of QCDSR to finite temperatures. In the QCDSR, at large distances
or low energies, the correlation function is formulated according
to hadronic parameters, called as the ``physical side'' or
``phenomenological side''. However, at short distances or high
energies, the correlator is defined with QCD parameters such as
quark masses and quark condensates. This side is named either the ``theoretical side'' or ``QCD side''. We can evaluate the
correlation function with these two sides, and there is a $q^2$ region
in which both sides can be equalized using the quark-hadron
duality hypothesis~\cite{Shifman}.

QCDSR method was first expanded to finite temperatures by Bochkarev and
Shaposhnikov~\cite{Bochkarev:1985ex}. In this version of the
QCDSR, analogous to vacuum sum rules the dual nature of the
correlator is employed. The features of hadrons in hot medium is identified by assuming both the operator product expansion
(OPE) and quark-hadron duality is valid, but the vacuum condensate values are displaced by their thermal versions.

To compute the mass and decay constants of the $\rho_2$, $\omega_2$, $\phi_2$ mesons, and their first excited states
within the TQCDSR approach, we start our calculation with the  temperature-dependent two-point correlation function as shown below:
\begin{eqnarray}\label{eq:correlation}
\mathrm{\Pi}_{\mu\nu,\alpha\beta}(q,T)&=&i \int d^{4}x e^{iq\cdot
(x-y)}\nonumber\\
&\times & Tr\left\{\varrho {\cal
T}\left[J_{\mu\nu}(x)J^\dagger_{\alpha\beta}(y)\right]\right\}_{y\rightarrow
0},
\end{eqnarray}
here $J_{\mu\nu}$ is the interpolating current belonging to the
$\rho^{(*)}_2$, $\omega^{(*)}_2$, $\phi^{(*)}_2$ mesons.  Here ${\cal T}$ is the time
ordered operator and the thermal density matrix is expressed with
\begin{equation}
\varrho=e^{- H/T}/Tr(e^{- H/T}),
\end{equation}
where $H$ is the QCD Hamiltonian and $T$ is the temperature of the
medium. The associated interpolating currents for the $\rho^{(*)}_2,~\omega^{(*)}_2$, $\phi^{(*)}_2$ states are given below
\cite{Aliev:2017apq}:
\begin{eqnarray}\label{eq:currentrho}
J _{\mu\nu}^{\rho^{(*)}_2}(x)&=&\frac{i}{2}\big[\bar u(x) \gamma_{\mu}
\gamma_{5} \olra{ \mathcal{\mathcal{D}}}_{\nu}(x) d(x) \nonumber\\
&+&\bar u(x)\gamma_{\nu} \gamma_{5} \olra{
\mathcal{\mathcal{D}}}_{\mu}(x) d(x)\big],
\end{eqnarray}
\begin{eqnarray}\label{eq:currentomega}
J_{\mu\nu}^{\omega^{(*)}_2}(x)&=&\frac{1}{2\sqrt{3}}\Bigg\{\Big[\overline{u}(x)\gamma_{\mu} \gamma_{5} \olra{\cal D}_{\nu}(x)u(x)\nonumber\\
&+&\overline{d}(x)\gamma_{\mu} \gamma_{5} \olra{\cal
D}_{\nu}(x)d(x)+\overline{s}(x) \nonumber\\ &\times&
\gamma_{\mu} \gamma_{5}\olra{\cal D}_{\nu}(x) s(x)\Big] +[\mu \leftrightarrow \nu ]\Bigg\},
\end{eqnarray}
\begin{eqnarray}\label{eq:currentfi}
J_{\mu\nu}^{\phi^{(*)}_2}(x)&=&\frac{1}{2\sqrt{6}}\Bigg\{\Big[\overline{u}(x)\gamma_{\mu} \gamma_{5} \olra{\cal D}_{\nu}(x)u(x)\nonumber\\
&+&\overline{d}(x)\gamma_{\mu} \gamma_{5} \olra{\cal
    D}_{\nu}(x)d(x)-2\overline{s}(x) \nonumber\\ &\times&
\gamma_{\mu} \gamma_{5}\olra{\cal D}_{\nu}(x) s(x)\Big] +[\mu \leftrightarrow \nu ]\Bigg\}.
\end{eqnarray}
In Eqs. (\ref{eq:currentrho}-\ref{eq:currentfi}), $\olra{\cal D}_{\mu}(x)$ shows the
derivative with respect to four-$x$ simultaneously acting on left
and right and it is described with
\begin{equation}\label{eq:covderivative}
\olra{\cal D}_{\mu}(x)=\frac{1}{2}[\overrightarrow{\cal
D}_{\mu}(x)- \overleftarrow{\cal D}_{\mu}(x)],
\end{equation}
\begin{eqnarray}\label{eq:covderivative2}
\overrightarrow{\cal D}_{\mu}(x)=\overrightarrow{\partial}_{\mu}+\frac{i}{2}g \lambda^a G_{\mu}^a,\nonumber\\
\overleftarrow{\cal
D}_{\mu}(x)=\overleftarrow{\partial}_{\mu}-\frac{i}{2} g \lambda^a
G_{\mu}^a,
\end{eqnarray}
where $\lambda^a$ ($a=1,8$) are the Gell-Mann matrices and
$G^a_\mu(x)$ are gluon fields. First, we focus on the ``\textit{physical side}'' of
the correlation function. In this side, i.e. at the hadron level, a complete set
of intermediate physical states with the same quantum numbers are embedded into
Eq.~(\ref{eq:correlation}) and then relevant integrals over four-$x$
are performed. Representing the axial-tensor mesons
with $A$ and their first excited states with $A^*$, the correlation
function can be written by
matrix elements of interpolating currents (for similar works see \cite{Agaev:2017tzv,Agaev:2017jyt,Agaev:2017lip})
\begin{eqnarray}\label{eq:phen1}
\mathrm{\Pi}
^{phys}_{\mu\nu,\alpha\beta}(q,T)&=&\frac{{\langle}\mathrm{\Omega}\mid
J_{\mu\nu}(0) \mid {A} \rangle \langle {A} \mid \bar
J_{\alpha\beta}(0)\mid \mathrm{\Omega}\rangle}{m_{A}^2(T)-q^2}\nonumber\\
&+& \frac{{\langle}\mathrm{\Omega}\mid J_{\mu\nu}(0) \mid {A^{*}} \rangle \langle {A^{*}} \mid \bar J_{\alpha\beta}(0)\mid \mathrm{\Omega}\rangle}{m_{A^{*}}^2(T)-q^2}\nonumber\\
&+& ...,
\end{eqnarray}
where $\mathrm{\Omega}$ indicates the hot medium and dots show the
contributions originating from the other excited states and continuum. The
matrix element $\langle \mathrm{\Omega} \mid J_{\mu\nu}(0)\mid
{A^{(*)}}\rangle$  and $\langle A^{(*)} \mid \bar{J}_{\alpha\beta}(0)\mid \mathrm{\Omega}\rangle$ is
 defined depending on the decay constant $f_{A^{(*)}}$ and
the mass $m_{A^{(*)}}$ in the following form
\begin{eqnarray}\label{eq:matrixelement}
\langle \mathrm{\Omega} \mid J_{\mu\nu}(0)\mid A^{(*)}\rangle=f_{A^{(*)}}(T)
m_{A^{(*)}}^3(T)~\varepsilon_{\mu\nu},
\end{eqnarray}
\begin{eqnarray}\label{eq:matrixelement2}
\langle A^{(*)} \mid \bar{J}_{\alpha\beta}(0)\mid \mathrm{\Omega}\rangle=f_{A^{(*)}}(T)
m_{A^{(*)}}^3(T)~\varepsilon^{'}_{\mu\nu},
\end{eqnarray}
here $\varepsilon_{\mu\nu}$ represents the polarization tensor
and the below relationship is valid:
\begin{eqnarray}\label{eq:polarizationt1}
\varepsilon_{\mu\nu}\varepsilon^{'}_{\alpha\beta}=\frac{1}{2}\eta_{\mu\alpha}\eta_{\nu\beta}+
\frac{1}{2} \eta_{\mu\beta}\eta_{\nu\alpha}-\frac{1}{3}\eta_{\mu\nu}\eta_{\alpha\beta},
\end{eqnarray}
where
\begin{equation}\label{eq:polarizationt2}
\eta_{\mu\nu}=-g_{\mu\nu}+\frac{q_\mu q_\nu}{m_{A^{(*)}}^2}.
\end{equation}
Inserting Eqs.~(\ref{eq:matrixelement}-\ref{eq:polarizationt2}) into Eq.~(\ref{eq:phen1}), the final
expression for the correlator belonging to the physical side is
obtained as
\begin{eqnarray}\label{eq:phen2}
&&\mathrm{\Pi}^{phys}_{\mu\nu,\alpha\beta}(q,T)=\Bigg[\frac{f_{A}^{2}(T)
m_{A}^{6}(T)} {m_{A}^2(T)-q^2}+\frac{f_{A^{*}}^2(T) m_{A^{*}}^6(T)} {m_{A^{*}}^2(T)-q^2}\Bigg]\nonumber\\
&&\times{\frac{1}{2}(g_{\mu\alpha}g_{\nu\beta}+g_{\mu\beta}g_{\nu\alpha})}+\mbox{other structures}.
\end{eqnarray}
Secondly, we compute the correlation function for ``\textit{QCD side}'' up to
certain order in the OPE expansion to get
thermal properties of the considered mesons. In this step, we can distinguish
the perturbative $\mathrm{\Gamma}(q^2,T)$ and non-perturbative
$\widetilde{\mathrm{\Gamma}}(q^2,T)$ contribution of the
correlation function in Eq.~(\ref{eq:correlation}):
\begin{equation}
\mathrm{\Pi}^{\mathrm{QCD}}(q^2,T)=\mathrm{\Gamma}(q^2,T)+\widetilde{\mathrm{\Gamma}}(q^2,T).~~~~~~~~
\end{equation}
At the quark level, i.e. in the QCD side, the correlation function can be defined in
the form of a dispersion relation:
\begin{equation}\label{eq:HadronSide}
\mathrm{\Gamma}(q,T) =\int \frac{\rho(s)}{s-q^2}~ds+subtracted~terms,
\end{equation}
here $\rho(s)$ is the spectral density function and expressed as:
\begin{eqnarray}\label{eq:HadronSide2}
\rho(s)&\equiv& \sum_n \delta(s-m_n^2)\langle \mathrm{\Omega} |J|n
\rangle \langle n
|J^\dagger |\mathrm{\Omega} \rangle \nonumber\\
&=& f_A^2 m_A^6 \delta(s -m_A^2) + f_{A^{*}}^2 m_{A^*}^6 \delta(s -m_{A^{*}}^2) \nonumber\\
&+& higher~states.
\end{eqnarray}
For computing the QCD side, the explicit expressions of the interpolating currents in Eqs.~(\ref{eq:currentrho}-\ref{eq:currentfi}) are embedded into
Eq.~(\ref{eq:correlation}). Then following standard manipulations, the QCD side of the correlation function is obtained as follows:
\begin{eqnarray}\label{eq:qcdsideRho}
&&\mathrm{\Pi}_{\mu\nu,\alpha\beta}^{ \rho^{(*)}_2}(q,T)=\frac{3i}{16}\int
d^{4}xe^{iq\cdot(x-y)}\Bigg\{Tr\Bigg[-\overrightarrow{\cal D}_{\beta}(y)\nonumber\\
&&\times S_d(y-x)\gamma_\mu\gamma_5 \overrightarrow{\cal D}_{\nu}(x)  S_u(x-y)\gamma_\alpha\gamma_5+S_d(y-x)\nonumber\\
&&\times \gamma_\mu\gamma_5 \overrightarrow{\cal D}_{\nu}(x)\overrightarrow{\cal D}_{\beta}(y)S_u(x-y)\gamma_\alpha\gamma_5+\overrightarrow{\cal D}_{\beta}(y)\overrightarrow{\cal D}_{\nu}(x)\nonumber\\
&&\times S_d(y-x)\gamma_\mu\gamma_5 S_u(x-y)\gamma_\alpha\gamma_5-\overrightarrow{\cal D}_{\nu}(x)S_d(y-x) \nonumber\\
&&\times \gamma_\mu\gamma_5 \overrightarrow{\cal D}_{\beta}(y)  S_u(x-y)\gamma_\alpha\gamma_5\Bigg]+\left[\beta\leftrightarrow\alpha\right]+\left[\nu\leftrightarrow\mu\right] \nonumber\\
&&+\left[\beta\leftrightarrow\alpha,
\nu\leftrightarrow\mu\right]\Bigg\}_{y\rightarrow 0},
\end{eqnarray}
\begin{eqnarray}\label{eq:qcdsideOmega}
&&\mathrm{\Pi}_{\mu\nu,\alpha\beta}^{ \omega^{(*)}_2}(q,T)=\frac{i}{16}\int
d^{4}xe^{iq\cdot(x-y)}\Bigg\{Tr\Bigg[\bigg(-\overrightarrow{\cal D}_{\beta}(y)\nonumber\\
&&\times S_u(y-x)\gamma_\mu\gamma_5 \overrightarrow{\cal D}_{\nu}(x)  S_u(x-y)\gamma_\alpha\gamma_5+S_u(y-x)\nonumber\\
&&\times \gamma_\mu\gamma_5 \overrightarrow{\cal    D}_{\nu}(x)\overrightarrow{\cal D}_{\beta}(y)S_u(x-y)\gamma_\alpha\gamma_5+\overrightarrow{\cal D}_{\beta}(y)\overrightarrow{\cal D}_{\nu}(x)\nonumber\\
&&\times S_u(y-x)\gamma_\mu\gamma_5 S_u(x-y)\gamma_\alpha\gamma_5-\overrightarrow{\cal D}_{\nu}(x)S_u(y-x)\nonumber\\
&&\times
\gamma_\mu\gamma_5 \overrightarrow{\cal D}_{\beta}(y) S_u(x-y)\gamma_\alpha\gamma_5\bigg)+\left(\beta\leftrightarrow\alpha\right)
+\left(\nu\leftrightarrow\mu\right) \nonumber\\
&&+\left(\beta\leftrightarrow\alpha,
\nu\leftrightarrow\mu\right)\Bigg]+[u \rightarrow d]+[u \rightarrow s]\Bigg\}_{y\rightarrow 0},
\end{eqnarray}
\begin{eqnarray}\label{eq:qcdsidefi}
&&\mathrm{\Pi}_{\mu\nu,\alpha\beta}^{ \phi^{(*)}_2}(q,T)=\frac{i}{32}\int
d^{4}xe^{iq\cdot(x-y)}\Bigg\{Tr\Bigg[\bigg(-\overrightarrow{\cal D}_{\beta}(y)\nonumber\\
&&\times S_u(y-x)\gamma_\mu\gamma_5 \overrightarrow{\cal D}_{\nu}(x)  S_u(x-y)\gamma_\alpha\gamma_5+S_u(y-x)\nonumber\\
&&\times \gamma_\mu\gamma_5 \overrightarrow{\cal    D}_{\nu}(x)\overrightarrow{\cal D}_{\beta}(y)S_u(x-y)\gamma_\alpha\gamma_5+\overrightarrow{\cal D}_{\beta}(y)\overrightarrow{\cal D}_{\nu}(x)\nonumber\\
&&\times S_u(y-x)\gamma_\mu\gamma_5 S_u(x-y)\gamma_\alpha\gamma_5-\overrightarrow{\cal D}_{\nu}(x)S_u(y-x)\nonumber\\
&&\times
\gamma_\mu\gamma_5 \overrightarrow{\cal D}_{\beta}(y) S_u(x-y)\gamma_\alpha\gamma_5\bigg)+\left(\beta\leftrightarrow\alpha\right)
+\left(\nu\leftrightarrow\mu\right) \nonumber\\
&&+\left(\beta\leftrightarrow\alpha,
\nu\leftrightarrow\mu\right)\Bigg]+[u \rightarrow d]+4[u \rightarrow s]\Bigg\}_{y\rightarrow 0}.
\end{eqnarray}
We replace the thermal light quark propagator
$S_{q}(x-y)$ in coordinate space in Eqs.~(\ref{eq:qcdsideRho}-\ref{eq:qcdsidefi}) defined in the form below:
\begin{eqnarray}\label{eq:lightquarkpropagator}
&&S_{q}^{ij}(x-y)= i\frac{\!\not\!{x}-\!\not\!{y}}{
    2\pi^2(x-y)^4}\delta_{ij}
-\frac{m_q}{4\pi^2(x-y)^2}\delta_{ij}\nonumber\\
&&-\frac{\langle \bar{q}q\rangle_T}{12}\delta_{ij}
-\frac{(x-y)^{2}}{192} m_{0}^{2} \langle
\bar{q}q\rangle_T\Big[1-i\frac{m_q}{6}(\!\not\!{x}-\!\not\!{y})\Big]\delta_{ij}
\nonumber\\
&&+\frac{i}{3}\Big[(\!\not\!{x}-\!\not\!{y})\Big(\frac{m_q}{16}\langle
\bar{q}q\rangle_T-\frac{1}{12}\langle
u^\mu\mathrm{\Theta}^{f}_{\mu\nu}u^\nu \rangle\Big)\nonumber\\
&&+\frac{1}{3}\Big(u\cdot(x-y)  \!\not\!{u}\langle
u^\mu\mathrm{\Theta}^{f}_{\mu\nu} u^\nu
\rangle\Big)\Big]\delta_{ij}
-\frac{ig_s G_{\mu\nu}}{32\pi^{2}(x-y)^{2}}\nonumber\\
&&\times \Big((\!\not\!{x}-\!\not\!{y})\sigma^{\mu\nu}
+\sigma^{\mu\nu}(\!\not\!{x}-\!\not\!{y})\Big)\delta_{ij},
\end{eqnarray}
where  $\mathrm{\Theta}^{f}_{\mu\nu}$ and $u_{\mu}$  are the fermionic
part of the energy momentum tensor and the four-velocity of hot medium, respectively. The temperature-dependent
 quark condensate is expressed in connection with vacuum condensate in the rest frame
$u_{\mu}=(1,0,0,0)$, $u^2=1$ \cite{Mallik:1997kj}.

After some long and standard calculations, correlation function of the QCD side is written with respect to the selected Lorentz
 structures just as in the physical side in Eq.~(\ref{eq:phen2}):
\begin{eqnarray}\label{eq:piqcd}
\mathrm{\Pi}^{\mathrm{QCD}}_{\mu\nu,\alpha\beta}(q^2,T)&=&\mathrm{\Pi}^{\mathrm{QCD}}(q^2,T)\left\{\frac{1}{2}%
(g_{\mu\alpha}g_{\nu\beta}+g_{\mu\beta}g_{\nu\alpha})\right\}
\nonumber \\
&+& \mbox{other structures}.
\end{eqnarray}
Next we obtain the correlation functions for both the physical and QCD sides separating the terms according to their structures.
Then, we need to eliminate the highest order particles from the lowest hadronic states. To do this taking derivative of unknown
polynomials in terms of $q^2$ in the correlators of both sides  based on the idea of QCDSR and employing the quark-hadron duality assumption, the following equality can be written:
\begin{equation}\label{eq:duality}
\mathcal{\widehat{B}}\mathrm{\Pi}^{phys}(q^2,T)=\mathcal{\widehat{B}}\mathrm{\Pi}^{\mathrm{QCD}}(q^2,T),
\end{equation}
here $\hat {\cal B}$ symbolizes the Borel transformation  defined by the undermentioned expression in which
$ \mathrm{F(x)} $ represents a function:
\begin{equation}\label{Borel}
\mathcal{\hat B}_{(q^2)}[\mathrm{F(x)}]\equiv \lim\limits_{\substack{n\rightarrow \infty \\ q^2=n M^2}} \frac{(-q^2)^{n}}{(n-1)!}\Bigg(\frac{d^n} {{d} q^{2n}}\Bigg) [\mathrm{F(x)}].
\end{equation}
After calculating the correlator belonging to the QCD and physical sides, equating the
 coefficients of selected structures $\left\{\frac{1}{2}(g_{\mu\alpha}g_{\nu\beta}+g_{\mu\beta}g_{\nu\alpha})\right\}$
and taking into account Borel transformation and quark-hadron duality,  we obtain the ground-state decay constant sum rule for
$\rho_2,~\omega_2$ and $\phi_2$ states as
\begin{eqnarray}\label{eq:fsumrule}
f_A^{2}(T)&=&\Bigg[\int_{s_{min}}^{s_{0}(T)}ds~\rho^{\mathrm{pert}}(s)~e^{-s/M^2}+\mathcal{\widehat{B}}\mathrm{\widetilde{\Gamma}}(q^2,T)\Bigg]\nonumber\\&\times& m_A^{-6}(T)~e^{m_A^2/M^2},
\end{eqnarray}
here $\mathrm{\widetilde{\Gamma}}$ represents the contribution of nonperturbative part belonging to the chosen structure.
 We have two expressions and two unknown parameters. One can extract the mass sum rule from Eq.~(\ref{eq:fsumrule}) easily performing
  derivative in terms of $(-1/M^2)$ where $M^2$ is the Borel mass parameter. So we also get the mass sum rule for the ground-state
  $\rho_2,~\omega_2$ and $\phi_2$ as
\begin{eqnarray}\label{eq:masssumrule}
&&~~~~~~~~~~~~~~~~~~~~~~~~~~m_A^{2}(T)=\frac{A_1}{B_1},\nonumber\\
&&A_1=\int_{s_{min}}^{s_{0}(T)}ds~\rho^{\mathrm{pert}}
(s)~s~e^{-s/M^{2}}+\frac{d}{d(-1/M^2)}{\mathcal{\widehat{B}}}\mathrm{\widetilde{\Gamma}}(q^2,T),\nonumber\\
&&B_1=\int_{s_{min}}^{s_{0}(T)}ds~\rho^{\mathrm{pert}}(s)~e^{-s/M^{2}}+\mathcal{\widehat{B}}\mathrm{\widetilde{\Gamma}}(q^2,T)
\end{eqnarray}
where $\sqrt{s_{min}}$ is the sum of quark contents of the
related mesons. As for the excited states sum rules of the examined axial-tensor mesons we get:
\begin{eqnarray}\label{eq:DecConSRExcited}
&&f_{A^*}^{2}(T)=\frac{1}{m_{A^*}^{6}}\Bigg[
\int_{s_{min}^*}^{s_{0}^{\ast}(T)}ds~\rho^{\mathrm{pert}}(s)~e^{(m_{A^*}^{2}-s)/M^{2}}
\nonumber\\
&&+e^{m_{A^*}^{2}/M^{2}}\mathcal{\widehat{B}}\mathrm{\widetilde{\Gamma}}(q^2,T)
-f_{A}^{2}m_{A}^{6}~e^{(m_{A^*}^{2}-m_{A}^{2})/M^{2}}\Bigg],\qquad
\end{eqnarray}
\begin{eqnarray}\label{eq:MassSRExcited}
&&~~~~~~~~~~~~~~~~~~~~~~~m^{2}_{A^*}(T)=\frac{A_2}{B_2},\nonumber\\
&&A_2=\int_{s_{min}^*}^{s_{0}^{\ast}(T)}ds~\rho^{\mathrm{pert}}(s)~s~e^{-s/M^{2}}-f_{A}^{2}m_{A}^{8}e^{-m_{A}^{2}/M^{2}},\nonumber\\
&&B_2=\int_{s^{\ast}_{min}}^{s_{0}^{\ast}(T)}ds~\rho^{\mathrm{pert}}(s)~e^{-s/M^{2}}+\mathcal{\widehat{B}}\mathrm{\widetilde{\Gamma}}(q^2,T)\nonumber\\
&&~~~~~~~~~~~~~~~~-f_{A}^{2}m_{A}^{6}e^{-m_{A}^{2}/M^{2}},
\end{eqnarray}
here $s_{0}^*(T)$ is the thermal continuum threshold parameter, which
separates the contribution of
``$A+A^*$'' from the ``higher resonances and
continuum''. Meanwhile sum rules depend on the same spectral
density $\rho ^{\mathrm{QCD}}(s)$ and the cut-off parameter must follow $s_0 < s_0^*$ where $s_0(0)$ and  $s_0^*(0)$ are the
 vacuum values of the continuum thresholds for the related ground states and first excited states respectively. As is mentioned above the mass
and decay constants of the ground state axial-tensor mesons enter into
Eqs.~(\ref{eq:fsumrule}-\ref{eq:MassSRExcited}) as input
parameters.

The spectral densities are parameterized as
\begin{eqnarray}
\rho(s)_{cont}=\rho^{\mathrm{QCD}}(s)\mathrm{\Theta}\big(s-s_0(T)\big)
\end{eqnarray}
with a single sharp pole pointing out the ground state hadron, and in the above equation
$\rho(s)_{cont}$ is the spectral density function of the continuum. $s_0(T)$ is the thermal cut-off parameter described in terms of  $s_0(0)$ ~\cite{Dominguez:2016roi,Borsanyi:2010bp,Bhattacharya:2014ara}:
\begin{eqnarray}\label{eq.sOT}
\frac{s_0(T)}{s_0(0)}= \bigg[ \frac{\langle {q}q
\rangle_T}{\langle \bar{q}q \rangle_0}\bigg]^{2/3}.
\end{eqnarray}
Next we move to the numerical analysis section.
%
%%%%%%%%%%%%%%%%%%%%%%%%%%%%%%%%%%%%%%%%%%%%%%%%%%%%%%%%%%%%%%%%%%%%%%%%%%%%%
\section{Numerical Analysis}
%%%%%%%%%%%%%%%%%%%%%%%%%%%%%%%%%%%%%%%%%%%%%%%%%%%%%%%%%%%%%%%%%%%%%%%%%%%%%

In this section we present numerical values of
input parameters used in our calculations in order to analyze the obtained sum rules,~i.e. Eqs.~(\ref{eq:fsumrule}-\ref{eq:MassSRExcited}). For the quark and mixed condensates we used
$\langle \bar{q} g_s \sigma G q\rangle= m_{0}^{2}\langle \bar{q}q
\rangle$, where $ m_{0}^{2}=(0.8\pm0.2)$ GeV$^2$, $
\langle0|\overline{u}u|0\rangle =
\langle0|\overline{d}d|0\rangle=-(0.24\pm0.01)^3$ GeV$^3$, $
\langle0|\overline{s}s|0\rangle=-0.8(0.24\pm0.01)^3$ GeV$^3$
~\cite{Shifman,Reinders,Ioffe,Narison:2003td}. The vacuum condensates are
parameters that do not depend on particles under consideration.
Their numerical values are extracted once and
are applicable in all sum rules calculations. The masses of $ u $, $ d $
and $ s $ quarks can be found in Ref.~\cite{Zyla}. They
are equal to $m_u = (2.16^{+0.49}_{-0.26})~\mathrm{MeV}$,
$m_{d}=(4.67_{-0.17}^{+0.48})~\mathrm{MeV}$ and $
m_{s}=(93_{-5}^{+11})~\mathrm{MeV}$.

During the calculations normalized thermal quark
condensate  is used in Eq.~(\ref{eq:lightquarkpropagator}) fitting Lattice data from Ref.~
\cite{Gubler:2018ctz} as follows representing $ q $,  $ u $ or $ d $ quarks
\begin{eqnarray}\label{eq:qbarqT}
\frac{\langle\bar{q}q\rangle_{T}}{\langle 0| \bar{q}q |0\rangle}=
\mathrm{C_1} e^{aT}+\mathrm{C_2}
\end{eqnarray}
and for the $ s $ quark
\begin{eqnarray}\label{eq:sbarsT}
\frac{\langle\bar{s}s\rangle_{T}}{\langle 0| \bar{s}s |0\rangle}=
\mathrm{C_3} e^{bT}+\mathrm{C_4},
\end{eqnarray}
here $ a=\mathrm{0.040~MeV^{-1}}$, $b=\mathrm{0.516~MeV^{-1}}$, $\mathrm{C_1}$$=-6.534\times10^{-4}$, $\mathrm{C_2=1.015}$, $\mathrm{C_3=-2.169\times10^{-5}}$ and
$\mathrm{C_4}$$=1.002$ are coefficients of the fit function.

Note that in Ref.~\cite{Gubler:2018ctz} the temperature dependence of quark condensates are presented up to temperature
 $T=300~\mathrm{MeV} $. However we parameterize them up
to the $T_c=165~\mathrm{MeV}$,  which is treated as the pseudocritical temperature for the crossover phase transition at zero chemical potential
 \cite{Azizi:2019cmj}. Then the fermionic part of the energy density is parameterized as \cite{Azizi:2015ona}
\begin{eqnarray}\label{eq:tetamumu}
\langle u^\mu\mathrm{\Theta}^{f}_{\mu\nu}u^\nu\rangle_T=T^{4}e^{\big(\lambda_1 T^{2}-\lambda_2 T\big)}-\lambda_3 T^5,
\end{eqnarray}
where $ \lambda_1=113.867 ~\mathrm{GeV^{-2}} $, $ \lambda_2=12.190 ~\mathrm{GeV^{-1}} $ and $ \lambda_3=10.141~ \mathrm{GeV^{-1}} $.

To check the reliability of thermal sum rules obtained, we examine whether the hadronic parameters of the particles handled give vacuum values. Note that in Eqs.~(\ref{eq:fsumrule}-\ref{eq:MassSRExcited}), the mass and decay constants QCD sum rules rely on the Borel mass parameter. Thus we determine the intervals of Borel mass parameter $M^2$ and
continuum threshold $s_0$. Our results should be insensitive to their variations because they are not completely physical quantities.
Given these circumstances, we
used $ s_0^{(*)}=(m_{A^{(*)}}+0.5)^2 $ condition for the ground and first excited states of the related mesons so that OPE convergence is satisfied.
Besides these criteria, values of physical properties of mesons have to be stable according to small changes of $s_0$ and $M^2$ as well.

\begin{figure}[h]
\begin{center}
\includegraphics[width=0.49\textwidth]{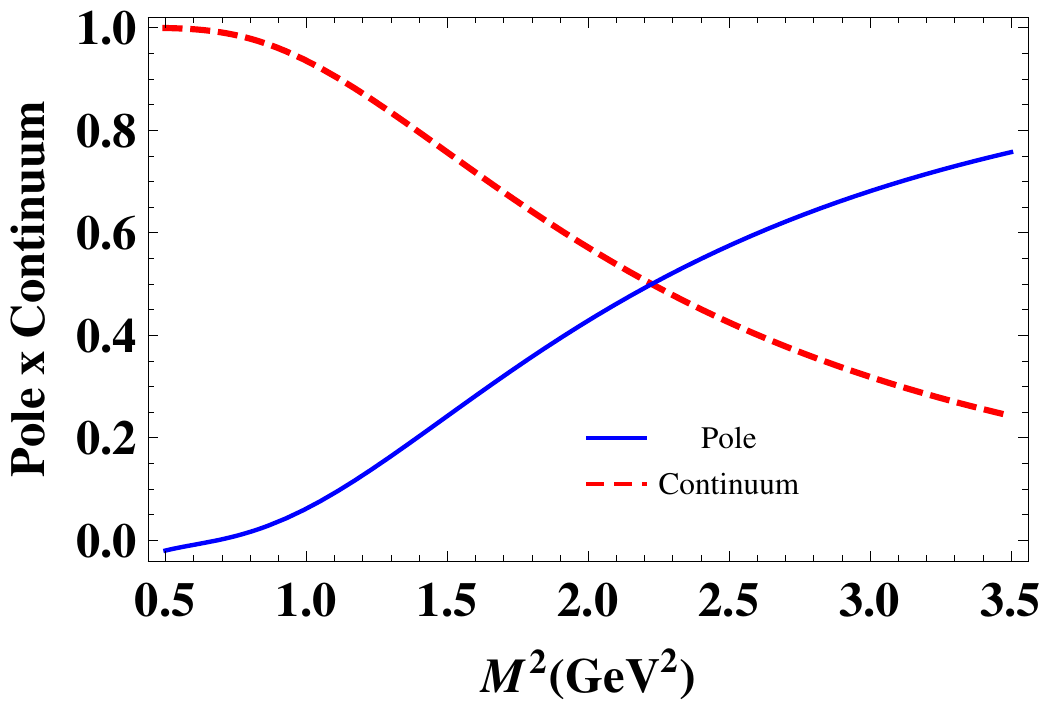}
\end{center}
\caption{Relative contributions of the pole~(red-dashed) and continuum~(blue) versus to the Borel parameter $M^2$ at
$s_0=5.95~\mathrm{GeV^2}$ for $\rho_2(1940)$ at $T=0$.\label{fig:Pole}}
\end{figure}

The gap of Borel mass parameter in QCD sum rule approach is determined by the following criteria:\\
\\
$\mathbf{a)}$ The lower bound of $M^2$ is fixed using the criterion of OPE convergence such that the contributions of
highest-dimensional operators are less than the $20\%$ of  total terms in OPE. In this computation this ratio is used as:
\begin{eqnarray*}\label{eq:convergence}
\frac{\mathrm{\Pi}^\mathrm{Dim5}}{\mathrm{\Pi}^{\rm all~terms}}<20\%
\end{eqnarray*}
where $\mathrm{\Pi}^{\mathrm{Dim5}}$ represents the contribution
from five dimensional operators. \\\\
$\mathbf{b)}$ For the upper bound of $M^2$ it is standard to employ the pole dominance condition which guarantees that the contribution of
continuum states is suppressed. One more condition for the intervals of these auxiliary parameters is the fact that since we extract information only from the ground state in the QCDSR approach, we have to ensure the pole contribution (PC) is larger than the continuum ones. To determine the PC in terms of $s_0$ and $M^2$ at $T=0$, we employ the below condition:\\
\begin{eqnarray*}\label{eq:ratio}
\mathrm{PC}=\frac{\mathrm{\Pi}(s_0,M^2, T=0)}{\mathrm{\Pi}(\infty,M^2,T=0 )}\geq50\%.
\end{eqnarray*}
Sum rule that do not obey above criteria is not applicable and must be discarded. Taking into account this condition we achieve
a $50\%$ pole contribution in the specified region and below present the graph in Figure \ref{fig:Pole}. \\ 
%%%%%%%%%%%%%%%%%%%%%%%%%%%%%%%%%%%%%%%%%%%%%%%%%%%%%%%%%%%%%%%%%%%%%%%%%%%%%%%
\\
We determine the values in Table \ref{table:MSQs0PDG} and \ref{table:MSQs0Regge} for the $s_0$ and $M^2$ parameters.\\
\begin{table}[h]
\caption{Borel and continuum threshold parameters working regions for the
$\rho_2^{(*)}$ and $\omega_2^{(*)}$ taking into account ``PDG'' data.}\label{table:MSQs0PDG}
\begin{center}
\begin{tabular}{|c|c|c|c|c|}
\hline\hline
Parameter               & $\rho_2$     & $\rho^*_2$    &  $\omega_2$&$\omega_2^*$ \\
\hline
$M^2(\mathrm{GeV^2})$   & $1.6-2.0 $   & $1.6-2.0 $    & $1.8-2.1$  & $1.8-2.1$  \\
$s_0^{(*)}(\mathrm{GeV^2})$   & $5.95 $      & $7.43 $       & $6.13$     & $7.26$ \\
\hline\hline
\end{tabular}
\end{center}
\end{table}
\begin{table}[h]
\caption{Borel and continuum threshold parameters working regions for the
$\rho_2^{(*)}$ and $\omega_2^{(*)}$ considering ``Regge Trajectory Model''.}\label{table:MSQs0Regge}
\begin{center}
\begin{tabular}{|c|c|c|c|c|}
\hline\hline
Parameter               & $\rho_2$  & $\rho^*_2$ & $\omega_2$ &$\omega_2^*$ \\
\hline
$M^2(\mathrm{GeV^2})$   & $1.6-1.8$   & $1.6-1.8$    &$1.3-1.5$     & $1.3-1.5$  \\
$s_0^{(*)}(\mathrm{GeV^2})$   & $4.82 $   & $5.95 $    & $4.82$     &  $6.67$  \\
\hline\hline
\end{tabular}
\end{center}
\end{table}\newline
Then using these numerical values we obtain the mass and decay constants of axial-tensor meson family and place the results in Table \ref{tab:MassResultsPDG}, \ref{tab:MassResultsRegge} and \ref{tab:DecConResults}. 
\begin{table}[h!]
	\caption{Mass values of the axial-tensor mesons at $T=0$ and comparison of the numerical values with experimental data form ``PDG''.}\label{tab:MassResultsPDG}
	\begin{tabular}{ccccc}
		\hline\hline
		&$m_{\rho_2}$&$m_{\rho^*_2}$& $m_{\omega_2}$&$m_{\omega^*_2}$ \\
		&($\mathrm{MeV}$)&($\mathrm{MeV}$)&($\mathrm{MeV}$)&($\mathrm{MeV}$)\\
		\hline
		Our&&&&\\
		Results&$1882$&$2258$&$1923$&$2288$\\
		Exp.\cite{Zyla}&$1940\pm40$&$2225\pm35$&$1975\pm20$&$2195\pm30$\\
		\hline\hline
	\end{tabular}
\end{table}\\
\begin{table}[h]
	\caption{Masses of the axial-tensor mesons at $T=0$ and comparison of the numerical values with the prediction of ``Regge Trajectory Model''.}\label{tab:MassResultsRegge}
	\begin{tabular}{ccccc}
		\hline\hline
		&$m_{\rho_2}$&$m_{\rho^*_2}$& $m_{\omega_2}$&$m_{\omega^*_2}$ \\
		&($\mathrm{MeV}$)&($\mathrm{MeV}$)&($\mathrm{MeV}$)&($\mathrm{MeV}$)\\
		\hline
		Our Results&$1604$&$1998$&$1668$&$1993$\\
		Regge Tr. Model~\cite{Guo:2019wpx}& $1696$&$1940$&$1696$&$1975$\\
		\hline\hline
	\end{tabular}
\end{table}\vline
\begin{table}[h]
	\begin{center}
		\caption{Decay constants of the axial-tensor mesons at $T=0$ and comparison of the numerical results with other theoretical predictions and experiment.}\label{tab:DecConResults}
		\begin{tabular}{ccccc}
			\hline\hline
			&$f_{\rho_2}$ & $f_{\rho^*_2}$ & $f_{\omega_2}$ & $f_{\omega^*_2}$\\
			\hline
			Our~~~~~~~~~~~~~~&&&&\\
			Results~$(\times10^{-2}) $& $6.96$& $3.32$& $5.98$& $1.85 $\\
			\hline
			QCDSR~\cite{Aliev:2017apq} $(\times10^{-2}) $& $7.4\pm0.1$& $-$& $6.2\pm0.4$& $-$\\
			\hline
			Exp.& $-$& $-$& $-$& $-$\\
			\hline\hline
		\end{tabular}
	\end{center}
\end{table}\\
These results are in good agreement with experiments and also the Regge Trajectory Model.   However this model claims that $\rho_2(1940)$ and $\omega_2(1975)$ mesons classified as ground states in PDG are indeed their excited states~\cite{Guo:2019wpx}. There are another two studies \cite{Barnes:1996ff,Godfrey:1998pd} denoting ground state masses of $\rho_2$ and $\omega_2$ as $ \sim1.7 $ GeV comparable with Regge Trajectory Theory~\cite{Guo:2019wpx}.  In this context we estimate
the mass and decay constant values of missing state $ \phi_2 $ in the $ J^{PC}=2^{--} $ nonet predicted by the Regge Trajectory Model;
\begin{eqnarray*}\label{eq:FiResults}
	m_{\phi_2}&=&1846~\mathrm{MeV},~~f_{\phi_2}=6.83\times10^{-2},\\
	m_{\phi_2^*}&=&2195~\mathrm{MeV},~~f_{\phi_2^*}=3.96
	\times10^{-2}
\end{eqnarray*}
in the Borel interval $ 1.1~ \mathrm{GeV^2}\leq M^2 \leq 1.3~\mathrm{GeV^2}$ and for the continuum thresholds $ s_0=5.77~ \mathrm{GeV^2} $,
$ s_0^*=7.02~\mathrm{GeV^2} $. Our result for the ground state mass of $\phi_2 $ resonance is consistent with the prediction in Ref.~\cite{Abreu:2020wio}
which finds the mass as $ m_{\phi_2}=1850~\mathrm{MeV} $ employing the Coulomb gauge Hamiltonian approach to QCD. It also agrees with the Ref.~\cite{Chen:2011qu} using Borel sum rules assuming contents of the related state as $ \bar{q}s $.

Finally, for all considered states, mass and decay constants versus $M^2$ and $s_0$ graphs are plotted at $T=0$ (but not presented in the paper
 for brevity) where dependencies of the hadronic parameters on $M^2$ and $s_0$ are shown to be weak. Therefore, we can say that the extracted sum rules
 are trustworthy in estimating the mass and decay constants of $\phi_2 $ and $ \phi_2^*$, and analyzing their thermal behaviors.
 Additionally, we draw the OPE convergence plot to ensure that the pole contribution is $50\% $ of the total contribution and determine the maximum value of $M^2$. For the $ \phi_2 $ and $ \phi_2^*$ resonances we need new precise experimental and also theoretical data to clarify the case. These missing mesons are still empirically unambiguous.
\begin{figure}[htbp]
\centering
\includegraphics[width=0.44\textwidth]{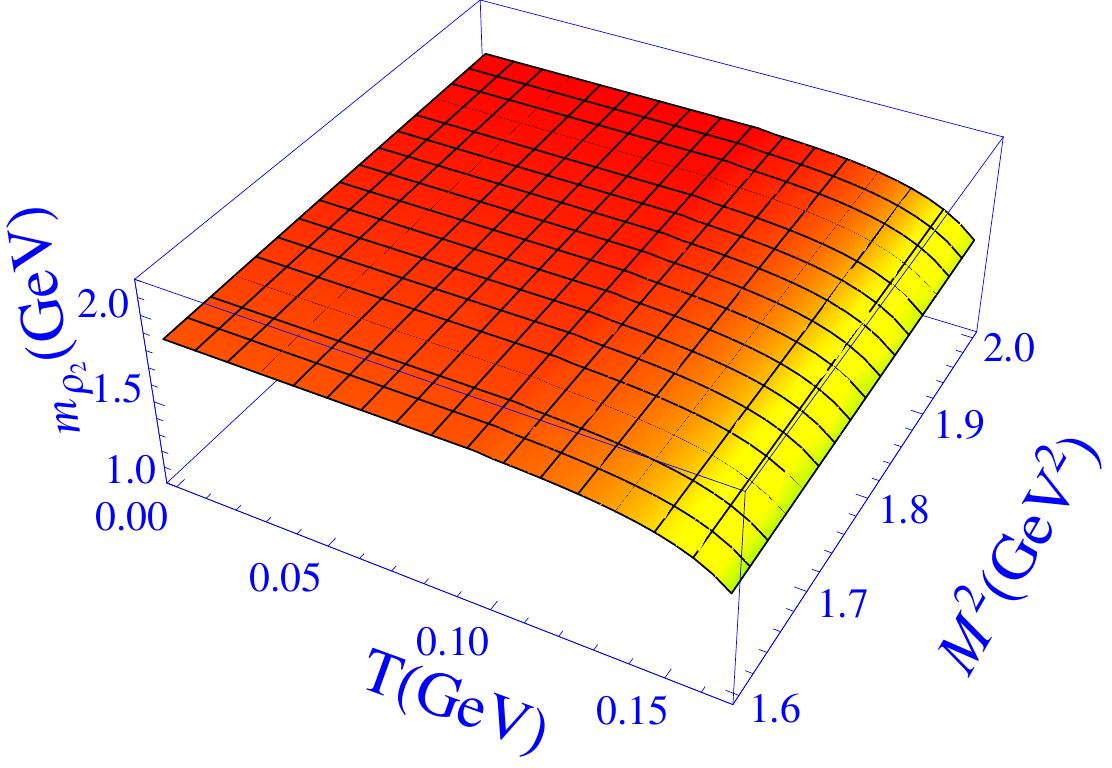}\vskip0.2cm
\includegraphics[width=0.44\textwidth]{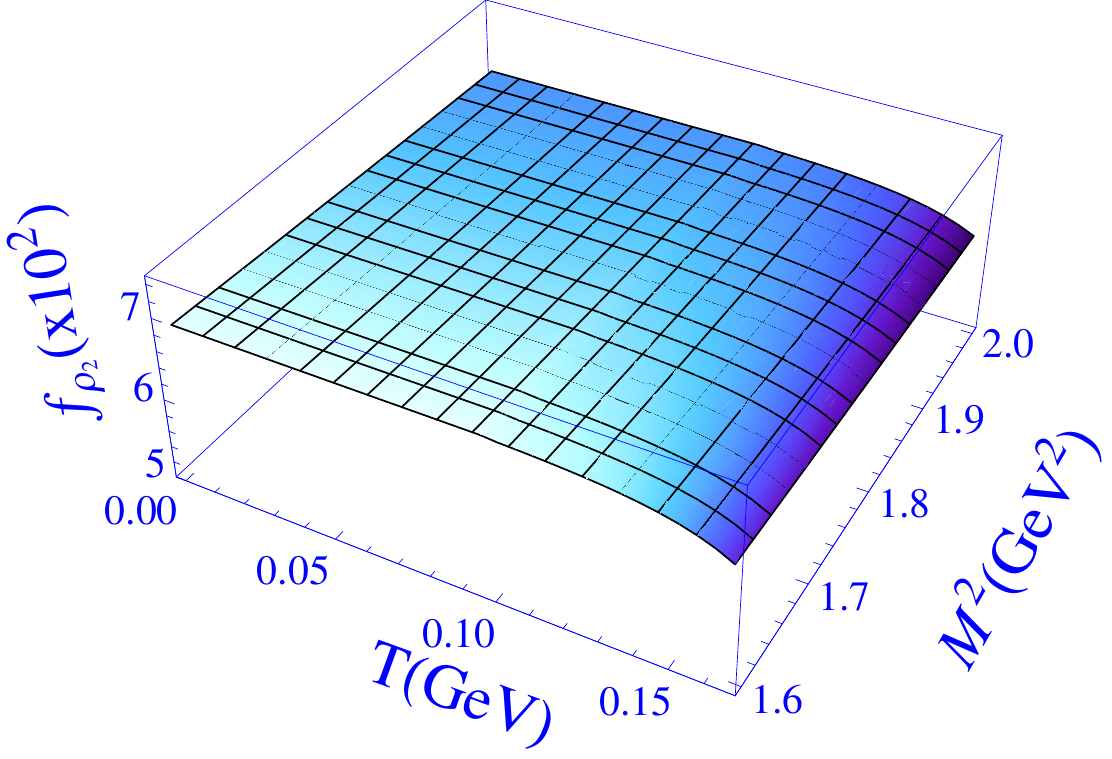}\vskip0.2cm
\includegraphics[width=0.44\textwidth]{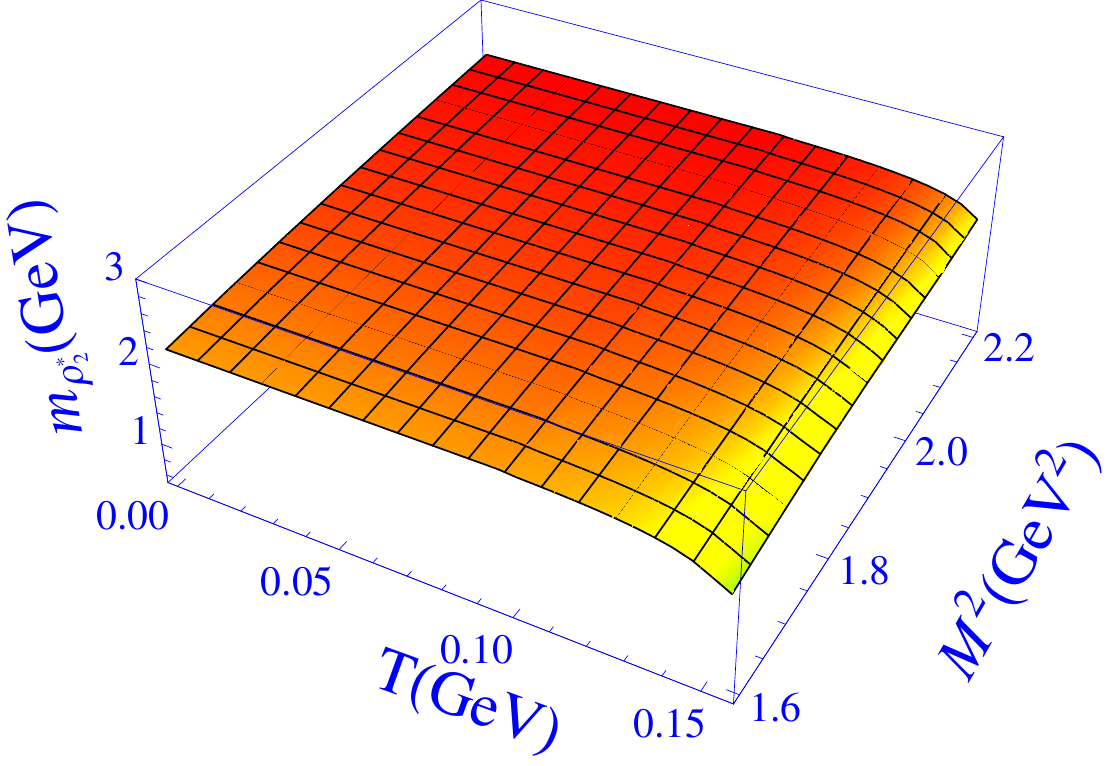}\vskip0.2cm
\includegraphics[width=0.44\textwidth]{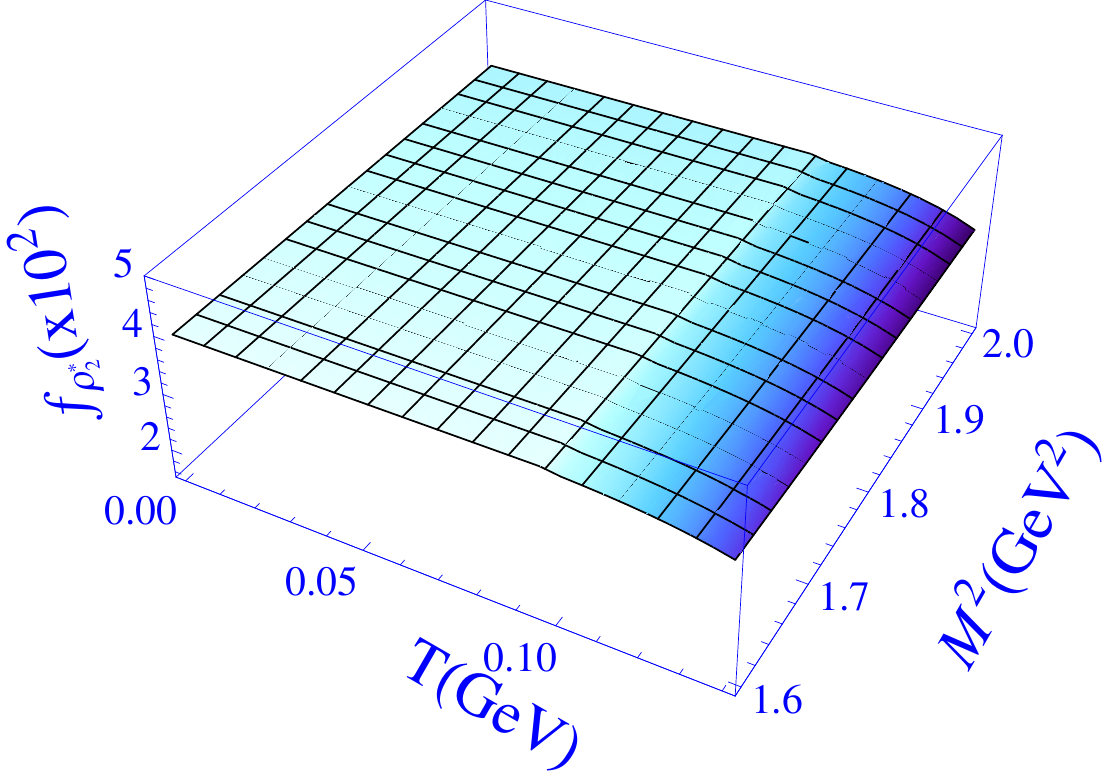}
\caption{The effect of temperature on mass~(first) and decay constant~(second) for $\rho_2$ meson and mass~(third) and decay constant~(fourth) for $\rho^*_2$ meson with respect to PDG data, respectively.\label{fig:RHOTPDG}}
\end{figure}
\section{Summary and Discussion}

In this article we have explored the hadronic properties of the $\rho^{(*)}_2,~\omega^{(*)}_2$ and $\phi^{(*)}_2$ mesons with  quantum numbers $J^{PC} = 2^{--}$ via the TQCDSR approach looking through the window of both Regge Trajectory Model and PDG data. Using the two-point thermal correlation function, we calculated the hadronic parameters of these particles up to dimension five. After obtaining the temperature dependence of mass and decay constant sum rules for the considered states, it is reduced to zero temperature to check the mass and decay constant values of $\rho^{(*)}_2,~\omega^{(*)}_2$ and $\phi^{(*)}_2$ at vacuum. To see the variations of the mass and decay constants in Eq.~(\ref{eq:fsumrule}-\ref{eq:MassSRExcited}) in terms of temperature, graphs are plotted for all considered mesons considering PDG data and also Regge Trajectory Model predictions by determining the related Borel mass and continuum threshold parameters separately. However, for the sake of brevity, we only present the 3-D mass graphs of $\rho_2$ and $\rho^*_2$ versus temperature and Borel mass according to PDG data in Figure~\ref{fig:RHOTPDG}.

Looking at analyses for the mass and decay constants of
$\rho_2^{(*)}$ and $\omega_2^{(*)}$, they remain unaffected until
$T\cong 0.12~\mathrm{GeV}$  with regard to both the PDG and  Regge
Trajectory Model data. Nevertheless after this temperature value
they start to deviate from vacuum values (For the rates of
change see the Table~\ref{table:perPDG} and \ref{table:perRegge}).
%%%%%%%%%%%%%%%%%%%%%%%%%%%%%%%%%%%%%%%%%%%%%%%%%%%%%%%%%%%%%%%%%%%
\begin{table}[h]
\caption{Percentage changes of mass and decay constants of the $\rho_2^{(*)}$ and $\omega_2^{(*)}$ compared with vacuum values in terms of
``PDG'' data at $ T_c=155~ \mathrm{MeV} $.}\label{table:perPDG}
\begin{center}
\begin{tabular}{|c|c|c|c|c|}
\hline\hline
Parameter           & $\rho_2$& $\rho^{*}_{2}$ &$\omega_2$ &$\omega_{2}^{*}$ \\
\hline
Mass (\%)           & 9       & 14         & 10        & 35 \\
Decay Constant (\%) & 4       & 1          & 3         & 14 \\
\hline\hline
\end{tabular}
\end{center}
\end{table}
\begin{table}[h]
\caption{Percentage variations of mass and decay constants of the $\rho_{2}^{(*)}$, $\omega_{2}^{(*)}$ and $\phi_{2}^{(*)}$ compared with vacuum
values according to ``Regge Trajectory Model'' data at $ T_c=155~ \mathrm{MeV} $.}\label{table:perRegge}
\begin{center}
\begin{tabular}{|c|c|c|c|c|c|c|}
\hline  \hline
Parameter & $ \rho_{2}$ & $ \rho_{2}^{*}$ &  $ \omega_{2}$&$ \omega_{2}^{*}$ &$ \phi_{2}$&$ \phi_{2}^{*}$\\
\hline\hline
Mass (\%)            & 10 & 26  & 9   & 19 & 10 & 25 \\
Decay Constant (\%)  & 3  & 34  & 2   & 18 & 2 & 14 \\
\hline\hline
\end{tabular}
\end{center}
\end{table}
%%%%%%%%%%%%%%%%%%%%%%%%%%%%%%%%%%%%%%%%%%%%%%%%%%%%%%%%%%%%%%%%%%%

As a result of these analyses, we conclude that the mass and decay
constants of $\rho^{(*)}_2,~\omega^{(*)}_2$ and $\phi^{(*)}_2$
mesons may dissociate at critical/pseudocritical temperature.
However we need more and precise experimental data to clarify the
situation. Although light mesons exist predominantly for a very
short time in heavy-ion collision experiments, we examine them for
more accurate interpretation of these experiments. To investigate
light unflavored mesons in extreme conditions is important to
understand the QCD vacuum, confinement and hadronisation phase of
the QGP and also whether mesons or baryons were formed earlier at
the initial stages of the universe. We hope that our numerical
results will be confirmed in near future both by experimental and
theoretical studies, and might help understand the nature of
strong interactions at finite temperatures.
\appendix
\section{Thermal spectral densities $\rho^{\mathrm{QCD}}(s, T)$
for the $\rho^{(*)}_2,~\omega^{(*)}_2$ and $\phi^{(*)}_2$ }\label{sec:App}
The spectral densities from QCDSR at high temperature
approximation is computed and presented explicitly in terms of
dimension in which contributions of the gluon condensates are
neglected due to its smallness ~\cite{Aliev:1981ju}.
The spectral density expressions for the $\rho^{(*)}_2,~\omega^{(*)}_2$ and $ \phi^{(*)}_2$ mesons up to dimension five are
found as follows:\\\\
-----
Perturbative Parts:
-----

\begin{eqnarray}\label{eq:rho}
\rho_{\rho^{(*)}_2}=\frac{3 s^2- 10 s m_u m_d }{80 \pi^2},
\end{eqnarray}

\begin{eqnarray}\label{eq:omega}
\rho_{\omega^{(*)}_2}=\frac{6s^2 - 5s (m_d^2 + m_u^2 + m_s^2)}{160
\pi^2},
\end{eqnarray}

\begin{eqnarray}\label{eq:fi}
\rho_{\phi^{(*)}_2}=\frac{12s^2 - 5s (m_d^2 + m_u^2 + 4 m_s^2)}{320
    \pi^2}.
\end{eqnarray}
----
Non-Perturbative Parts:
----

\begin{eqnarray}\label{eq:nonpertrho}
\widetilde{\mathrm{\Gamma}}_{\rho^{(*)}_2}(q,T)&=&\frac{4\langle
u\mathrm{\Theta^{f}}u\rangle(q\cdot
u)^{2}}{3q^{2}}\nonumber\\
&-&\frac{m_0^{2}(m_u\langle\bar{u}u\rangle + m_d
\langle\bar{d}d\rangle)}{4 q^{2}},\quad\quad\quad\quad\quad\quad\quad\quad
\end{eqnarray}

\begin{eqnarray}\label{eq:nonpertomega}
\widetilde{\mathrm{\Gamma}}_{\omega^{(*)}_2}(q,T)&=&\frac{4\langle
u\mathrm{\Theta^{f}}u\rangle(q\cdot u)^{2}}{9q^{2}}\nonumber\\
&-&\frac{41 m_0^{2}(m_d\langle\bar{d}d\rangle +
m_u\langle\bar{u}u\rangle + m_s\langle\bar{s}s\rangle)}{144
q^{2}},\quad\quad
\end{eqnarray}

\begin{eqnarray}\label{eq:nonpertfi}
\widetilde{\mathrm{\Gamma}}_{\phi^{(*)}_2}(q,T)&=&\frac{8\langle
    u\mathrm{\Theta^{f}}u\rangle(q\cdot u)^{2}}{27q^{2}}\nonumber\\
&-&\frac{41 m_0^{2}(m_d\langle\bar{d}d\rangle +
    m_u\langle\bar{u}u\rangle + 4 m_s\langle\bar{s}s\rangle)}{288
    q^{2}}.\quad\quad
\end{eqnarray}

\end{document}